# 3D printing of optical materials: an investigation of the microscopic properties

Luana Persano[a], Francesco Cardarelli[a], Arkadii Arinstein[b], Sureeporn Uttiya,[a] Eyal Zussman,[b] Dario Pisignano,[a,c] Andrea Camposeo*[a]

[a] NEST, Istituto Nanoscienze-CNR, Piazza S. Silvestro 12, I-56127 Pisa, Italy; [b]Department of Mechanical Engineering, Technion-Israel Institute of Technology, Haifa, 32000, Israel; [c]Dipartimento di Fisica, Università di Pisa, Largo B. Pontecorvo 3, I-56127 Pisa, Italy

**ABSTRACT**

3D printing technologies are currently enabling the fabrication of objects with complex architectures and tailored properties. In such framework, the production of 3D optical structures, which are typically based on optical transparent matrices, optionally doped with active molecular compounds and nanoparticles, is still limited by the poor uniformity of the printed structures. Both bulk inhomogeneities and surface roughness of the printed structures can negatively affect the propagation of light in 3D printed optical components. Here we investigate photopolymerization-based printing processes by laser confocal microscopy. The experimental method we developed allows the printing process to be investigated in-situ, with microscale spatial resolution, and in real-time. The modelling of the photo-polymerization kinetics allows the different polymerization regimes to be investigated and the influence of process variables to be rationalized. In addition, the origin of the factors limiting light propagation in printed materials are rationalized, with the aim of envisaging effective experimental strategies to improve optical properties of printed materials.

## 1. INTRODUCTION

The technologies for the fabrication of 3D objects are currently undergoing a fast and impressive evolution, enabling novel structures to be realized with a variety of materials, including metals [1], ceramics [2], semiconductors [3], nanocomposites [4] and soft materials [5]. The possibility of realizing objects with an unprecedented level of multiscale geometrical complexity has allowed novel physical effects to be explored, such as the negative thermal expansion [6] and the sign reversal of Hall coefficient [7]. Also, artificial objects with unconventional mechanical properties are being realized [8]. In this framework, photopolymerization processes are gaining a renewed interest, being at the base of 3D printing processes working through light exposure. In such methods, a 3D object is typically built in a layer-by-layer fashion by curing a liquid pre-polymer mixture through the exposure to UV light. This can be accomplished by using a UV laser, as in stereolithography (STL) [9], where a laser beam is focused in the pre-polymer bath, locally inducing the polymerization of the used monomers or oligomers upon addition of suitable photoinitiators. The individual layers forming the 3D objects are built by scanning the laser beam, typically by exploiting opto-mechanical systems. As a consequence, the spatial resolution of UV-based STL is limited by the capability of depositing thin layers of pre-polymers for the vertical direction, while diffraction effects determine the in-plane resolution. A laser beam can be focused to reach spot size down to 1 μm by exploiting optics with high numerical aperture (*NA*) and utilizing small printing areas (< 1 mm$^2$). In fact, in commercially available UV-based STL systems, which are designed for large-area printing ($10^2$-$10^3$ cm$^2$) and generally exploit low-*NA* optics, the spatial resolution is limited to tens of micrometers.

The spatial resolution can be significantly improved to values below the diffraction limit, by exploiting two-photon absorption processes [10]. Two-photon lithography is currently exploited for the fabrication of micro-optical systems [11] and of other highly precise micro-objects [12], even in continuous runs and in combination with microfluidic systems [13]. Another critical aspect for the development of next-generation 3D printing methods is the achievable throughput. The throughput of stereolithographic technologies, which are inherently serial processes, is limited by the scanning velocities of the laser beam, a drawback that has motivated the development of alternative, faster approaches.

*andrea.camposeo@cnr.it; phone +39 050 509517.

Among the others, digital light processing (DLP) overcomes some limitations of the STL by projecting individual layers in the liquid bath during a single exposure, through either digital micro-mirror array devices [14] or dynamic liquid-crystal masks [15]. While showing spatial resolution in the range 1-10 µm [15], DLP systems can reach printing speeds up to $10^3$ mm per hour, and almost continuous printing by exploiting the polymerization-inhibiting effect of oxygen [16]. In this method, oxygen is allowed to diffuse across the pre-polymer, forming a thin layer of uncured liquid material between the polymerized layer and the optical window used for light exposure, and enabling a continuous renewal of reactive, liquid resin in the layer exposed to UV light. In other approaches, 3D structures are realized by extrusion of a filament made of a photocurable ink through a robotically-controlled syringe; the ink is then polymerized by exposure to the UV light generated by a LED, within few seconds following extrusion [17].

Overall, exposure-based technologies have been employed for the production of various structures, including scaffolds for tissue engineering [18], lab-on-chip devices [19] and optical interconnects [20]. In particular, the fabrication of optical components by photopolymerization-based 3D printing demands for the development of processes leading to highly transparent and uniform printed components, to minimize optical losses associated with absorption and scattering of light through the objects. Research efforts are currently focused to print transparent materials, which led, for instance, to the recent demonstration of 3D-printed objects made by fused silica glass through a hybrid approach based on STL of a photocurable silica nanocomposite and post-printing, thermal conversion to high-quality glass [21]. These materials showed a transparency above 90 % in the visible spectral range (300-1,000 nm). Finally, the uniformity of the printed structures is also relevant. Local variations of photoinitiator concentrations and of the intensity of the curing light can generate non-uniform polymerization of the used pre-polymers, resulting in turn in local variations of the refractive index and higher scattering of photons transported in the components. The occurrence of such detrimental effects might be prevented by means of suitable methodologies allowing the polymerization kinetics to be monitored in-situ and in real-time during 3D printing processes.

Here, we report on the investigation of the photopolymerization kinetics by measuring the temporal evolution of the light backscattered by the sample, during the exposure of a photosensitive pre-polymer to a focused UV laser beam in a confocal laser scanning microscopy system. The material used in the study is SU8, an epoxy-based negative-tone photoresist which is widely used in microelectronics and micro-mechanics especially for the production of high-aspect ratio structures [22-24]. SU8 has been also employed in laser-based 3D printing processes, exploiting both linear and nonlinear absorption [25, 26], with minimum feature size down to 190 nm. The photopolymerization kinetics of SU8 has been studied by various methods [27,28], which provided the temporal evolution of the ensemble-averaged, macroscopic process. For instance, both catalytic and retarding effects have been highlighted for the polymerization kinetics by using nanomaterials (dielectric nanoparticles and graphene) embedded in SU8. Our approach allows the photopolymerization kinetics to be studied with micrometric spatial resolution and with time resolution of 0.1 s. The method here developed provides a tool for the microscopic investigation of the 3D printing processes of optical materials, that is particularly relevant for heterogeneous systems and nanocomposites, where the addition of guest molecules and particles to transparent matrices is expected to locally vary the polymerization process through geometrical constraints, interfacial effects, and photochemical reactions.

## 2. METHODS

The system used in this study was SU8 (MicroChem) deposited on cleaned quartz substrates (1×1 cm$^2$) by spin coating at 1000-3000 rpm. The thickness of the SU8 films (in the range 1-5 µm) was measured by a stylus profilometer (Dektak). To this aim, spin-coated SU8 films were first soft-baked in a thermal oven at ~95 °C for 1 hour and subsequently polymerized in a UV oven at ~40 °C for 2 hours (exposure wavelength 405 nm). After UV-exposure, SU8 films were post-baked at ~95 °C for 2 minutes for full curing. UV-visible transmission spectra were acquired by using a spectrophotometer, with spectral resolution of 0.5 nm.

The photopolymerization kinetics was characterized by confocal microscopy, by using an inverted optical microscope equipped with a confocal laser scanning head (Olympus, FV1000). SU8 films were exposed to a focused laser beam, emitted by a diode laser (emission wavelength, $\lambda_{em}$=405 nm), which was focused onto the sample through a 10× objective ($NA$=0.4, spot size about 1.2 µm). The incident laser power could be varied linearly in an interval 0.01-1.1 mW, with a maximum intensity of the order of $10^8$ mW cm$^{-2}$. The same laser was used to probe the photopolymerization kinetics, by collecting the light backscattered by the sample through the excitation objective. The intensity of the backscattered light was analyzed by using a photomultiplier detector, which provides a measure of the temporal evolution of the backscattered light signal, $I(t)$, with a time resolution of 0.1 s. The exposed regions had areas of about 6.25 μm². This method allowed the typical exposure conditions of UV STL processes to be reliably reproduced and investigated. After exposure, confocal micrographs of the exposed regions were collected by measuring the intensity of light backscattered by the sample, using the same experimental configuration utilized for the photopolymerization kinetic measurements, upon decreasing the intensity of the incident laser by an order of magnitude to avoid additional polymerization.

## 3. RESULTS AND DISCUSSION

Figure 1a shows the optical transmission spectra of SU8 samples before and after UV-induced polymerization. The material is transparent in the range 400-900 nm, with optical transmission >90 %. This property, in combination with a high mechanical modulus and refractive index of about 1.6, makes SU8 suitable for 3D optical components and interconnects. We exploited confocal laser scanning microscopy to probe the polymerization kinetics in experimental conditions typical of laser-based 3D printing processes. Figure 1b shows a sequence of confocal microscopy images of the regions exposed to the focused laser beam, with varying exposure times. While for exposure times < 100 s a negligible signal is measured, upon increasing the exposure time above 120 s a bright spot emerges in correspondence of the exposed area, with full width at half maximum (FWHM) of about 2.8 μm, almost constant for exposure times > 100 s (inset of Fig. 1c) and roughly matching the size of the exposed area (2.5 μm).

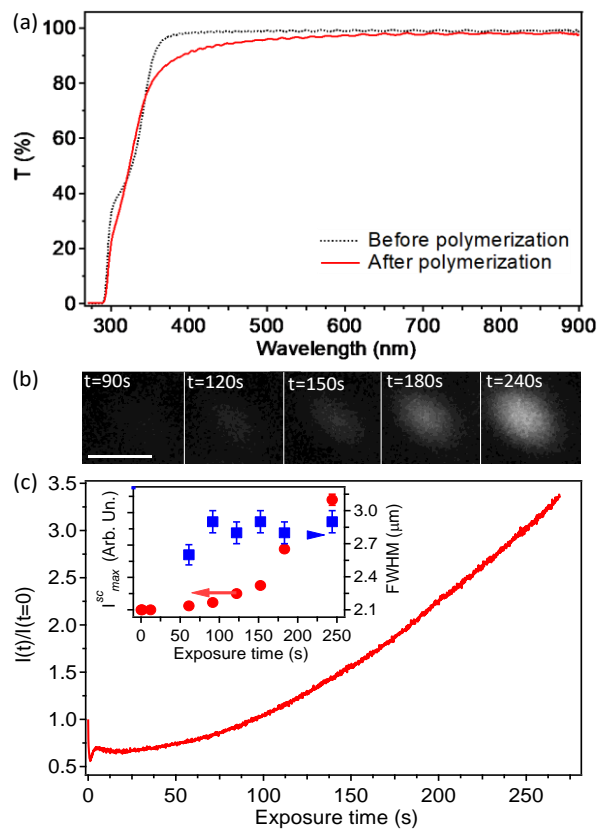

Figure 1. (a) Spectra of the light intensity transmitted (T) by a sample before exposure to UV light (black dotted line) and after UV-induced polymerization (red continuous line). Sample thickness: 4.5 μm. (b) Confocal maps of the intensity of the backscattered light after various exposure times, $t$. From left to right: $t$=90s, 120s, 150s 180s and 240 s. Scale bar: 5 μm. (c) Example of the temporal evolution of the intensity of the backscattered light, $I(t)$, collected in confocal configuration, normalized to the intensity at $t$=0. Inset: plot of the structure FWHM (blue squares, right vertical scale) and of the maximum

intensity of the backscattered light, $I^{sc}_{max}$ (red circles, left vertical scale), calculated from confocal maps of the backscattered light intensity (shown in panel b) for increasing exposure times. Exposure laser power: 1 mW.

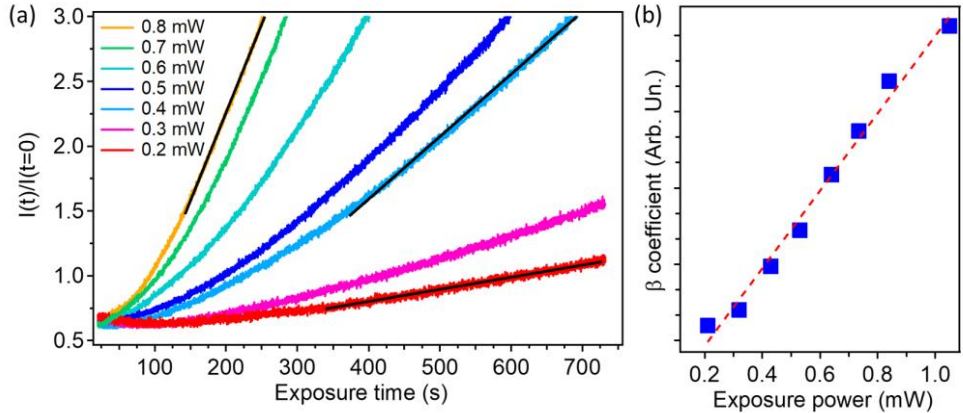

Figure 2. (a) Temporal evolution of the intensity of the backscattered light, $I(t)$, collected in confocal configuration and normalized to the intensity at $t=0$. Data acquired with different powers of the incident laser. Black continuous lines are examples of linear fits performed in the time interval where the cluster-cluster model is effective. (b) Dependence of the coefficient β, from Equation (8), of the linear fit on the power of the laser used for UV-photopolymerization. The red dashed line is a linear fit to the data.

The bright spot is due to the polymerized SU8, diffusing more light due to the variation of the refractive index of the organic material following curing (of the order of $10^{-3}$ [29]) compared to the surrounding, non-polymerized material. By increasing the exposure time, the intensity of the backscattered signal increases following a non-linear trend (Figure 1c). These data directly provide the measure of the temporal evolution of the photopolymerization process. Measurements performed both in different samples and in various regions of the samples at comparable exposure parameters did not show significant variation of the behavior of collected back-scattered intensity, with minor sample-to-sample variation attributable to local inhomogeneities of the sample thickness and different focusing conditions. The dependence of the photopolymerization kinetics on the power of the UV laser is shown in Figure 2a. The data clearly show how a slower polymerization is associated to the lower power of the incident printing laser.

The complex trend of the time evolution of the light intensity backscattered by UV-cured SU8 reflects the various stages of photopolymerization. SU8 curing is known to proceed through a first stage where protonic acids are produced by the interaction with UV light, followed by a second stage where cationic polymerization occurs [27,28]. An effective approach for describing the polymerization kinetics is to adopt for the cluster concept for linked regions formed during the polymerization, using the well-known mathematical model for cluster-cluster aggregation based on Smoluchowski equation system [30]. At least two stages can be envisaged during cluster growth: the first one is typically dominated by individual growing clusters, increasing in size by addition of elemental units (monomers or oligomers), whereas the second stage is characterized by cluster-cluster interaction and merging. For printing of optical materials this second stage is very important, because an effective coalescence of the various clusters would avoid the presence of unintentional spatial variations of refractive index and/or absorption that can enhance the diffusion of light propagating through the printed structure as mentioned above.

The cluster-cluster irreversible aggregation process can be described by the following equation system [31]:

$$\frac{\partial c_n(t)}{\partial t} = \frac{1}{2}\sum_{m<n} K_{m,n-m} c_m(t) c_{n-m}(t) - c_n(t) \sum_{m \geq 1} K_{n,m} c_m(t), \quad (1)$$

where $c_n$ is the density of $n$-sized clusters (containing $n$ elemental units), $K_{n,m}$ is the probability of $n$- and $m$-clusters to join into one $k$-sized cluster ($k = n + m$). In addition, the following condition has to be accounted due to the mass conservation law:

$$\sum_{n=1}^{\infty} nc_n(t) = M \qquad (2)$$

The equation system (1) can be approximated in the integral form as:

$$\frac{\partial c(x,t)}{\partial t} = \frac{1}{2}\int_0^k K(y,x-y)c(y)c(x-y)dy - c(x)\int_0^{\infty} K(x,y)c(y)dy, \qquad (3)$$

whereas the mass conservation law also can be written in the integral form:

$$\int_0^{\infty} yc(y,t)dy = M . \qquad (4)$$

Considering the simplest case of uniform probabilities $K(x,y) = K$, the following equation can be derived:

$$\frac{\partial c(k,t)}{\partial t} = \frac{K}{2}\int_0^k c(y)c(k-y)dy - Kc(k)\int_0^{\infty} c(y)dy, \qquad (5)$$

having the solution:

$$c(x,t) = Ma^2(t)\exp[-a(t)x] \qquad (6)$$

with $a(t) = 2/KMt$. In a simplified picture, the intensity of back-scattered light, $I(t)$ can be considered proportional to the mean size of the formed clusters:

$$I(t) = I_0 \times \kappa \bar{c}(t), \qquad (7)$$

$$\bar{c}(t) = \int_0^{\infty} xc(x,t)dx \bigg/ \int_0^{\infty} c(x,t)dx = M \bigg/ \int_0^{\infty} c(x,t)dx = \frac{KMt}{2} = \beta t \qquad (8)$$

which shows that in the cluster-cluster aggregation regime the intensity of the backscattered light is expected to vary linearly with exposure time, with a slope coefficient ($\beta$) proportional to the incident light intensity $I_0$. Indeed, a linear increase of $I(t)$ is observed at long exposure times for all the investigated incident powers (Figure 2a). Linear-fitting of the data in such regime highlights a linear dependence of the $\beta$ coefficient on printing laser power (Figure 2b), as expected on the base of the developed model.

A more complex behavior is observed in the first polymerization stage (Figure 3a). Here a rapid decrease of $I(t)$ is found, down to a minimum value reached at a time, $t_{min}$, in the range of 1-4 s, depending on the power of the UV laser. The dependence of $t_{min}$ and of the corresponding minimum value of $I(t)$ on the exposure power is shown in Figure 3b. An almost linear decrease of both the quantities is observed up to an incident laser power of about 0.7 mW, whereas they remain almost constant for values of the incident power in the range 0.8-1.1 mW. Here, various effects are expected to contribute and to affect the observed behavior, involving the absorption of incident light by photoinitiators and production of protonic acid, which triggers the cationic polymerization [27,28]. At this stage, local heating effects cannot be ruled out, which can in turn lead to local changes in refractive index and density. A full description of the phenomena occurring for exposure time < 10s require additional and extensive experimental investigations, that are currently underway in our laboratory.

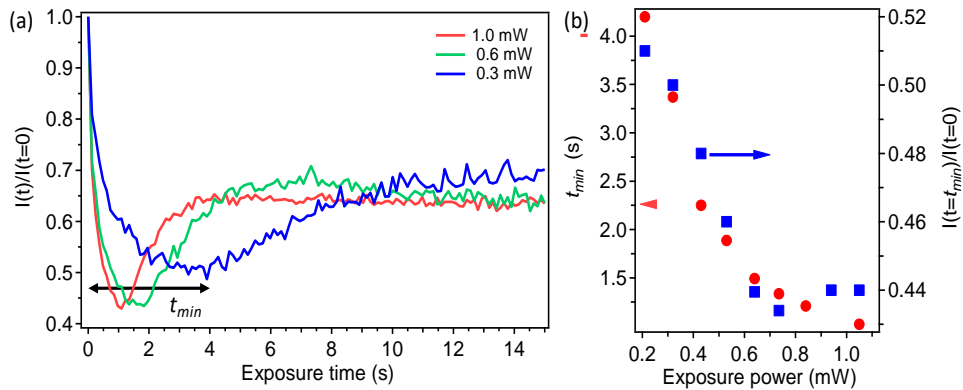

Figure 3. (a) Evolution of the intensity of the backscattered signal, $I(t)$, at short times. The signal is normalized to the intensity at $t$=0 and studied for incident laser power 1.0, 0.6 and 0.3 mW. (b) Dependence of $t_{min}$ (red circles, left vertical scale) and of $I(t=t_{min})/I(t=0)$ (blue squares, right vertical scale) on the power of the laser used for UV photopolymerization. For better clarity, $t_{min}$ is shown in panel (a) for data measured at 0.3 mW.

## 4. CONCLUSIONS

In summary, we introduced a novel methodology for real-time monitoring of photopolymerization processes, based on measuring the intensity of the printing laser that is back-scattered by the sample. The temporal evolution of the polymerization process can be studied in-situ, with micrometric spatial resolution, without the need of additional probes or additional light sources as reported in previous works [32]. The photopolymerization kinetics of the SU8 photoresist was investigated by the developed method, varying the intensity of the printing laser. The measured kinetics is well described by a cluster-cluster aggregation model at longer exposure times. The method here proposed will allow the polymerization phenomena to be locally monitored and controlled, opening a new way to highlight and suppress inhomogeneous cured features during 3D printing processes, and to ultimately improve the optical uniformity of the printed structures in terms of refractive index and transparency. Next steps will include the rationalization of the signal and kinetics measured at short exposure times, and the realization of 3D printed optical components with enhanced uniformity and transparency.

*Acknowledgments*. The research leading to these results has received funding from the European Research Council (ERC) under the European Union's Horizon 2020 research and innovation programme (grant agreement No. 682157, "xPRINT"). The authors also acknowledge L. Romano for film depositions.